%% file: main.tex
\documentclass[9pt,journal,compsoc]{IEEEtran}

\ifCLASSOPTIONcompsoc
  \usepackage[nocompress]{cite}
\else
  \usepackage{cite}
\fi
\ifCLASSINFOpdf
   \usepackage[pdftex]{graphicx}
   \usepackage{subcaption}
\else
\fi
\usepackage{amsmath,amssymb,amsfonts}
\usepackage{algorithmic}
\usepackage{graphicx}
\usepackage{textcomp}
\usepackage{xcolor}
\usepackage{fancyhdr}
\usepackage[english]{babel}
\usepackage{blindtext}
\usepackage{xspace}
\usepackage{tabularx}
\usepackage{multirow}
\usepackage{subcaption}
\usepackage{footnote}
\usepackage{color,soul}
\usepackage{blindtext}

\begin{document}
%
\title{Accelerating Graph Analytics on a Reconfigurable Architecture with a Data-Indirect Prefetcher
\vspace{-10pt}}
%
%
%
%
\vspace{-15pt}
\author{Yichen~Yang, Jingtao~Li, Nishil~Talati, Subhankar~Pal, Siying~Feng, Chaitali~Chakrabarti, Trevor~Mudge, Ronald~Dreslinski
\IEEEcompsocitemizethanks{\vspace{-5pt}\IEEEcompsocthanksitem Yichen Yang, Nishil Talati, Subhankar Pal, Siying Feng, Trevor Mudge and Ronald Dreslinski were with University of Michigan, Ann Arbor,
MI, 48105.
E-mail: yangych@umich.edu
\IEEEcompsocthanksitem Jingtao Li and Chaitali Chakrabarti are with Arizona State University, Tempe, AZ 85281.
\vspace{-7pt}}
\vspace{-7pt}}

\input{00_abstract}

\maketitle

\IEEEdisplaynontitleabstractindextext

%
\IEEEpeerreviewmaketitle

\input{01_introduction}
\input{02_background}
\input{03_arch}
\input{04_method}

\input{05_result}
\input{07_conclusion}







\vspace{-10pt}
\bibliographystyle{IEEEtran}
\bibliography{refs}
\end{document}

%% file: 00_abstract.tex
\IEEEtitleabstractindextext{%
\vspace{-7pt}
\begin{abstract}
The irregular nature of memory accesses of graph workloads makes their performance poor on modern computing platforms.
On manycore reconfigurable architectures (MRAs), in particular, even state-of-the-art graph prefetchers do not work well (only 3\% speedup), since they are designed for traditional CPUs.
This is because caches in MRAs are typically not large enough to host a large quantity of prefetched data, and many employs shared caches that such prefetchers simply do not support.
This paper studies the design of a data prefetcher for an MRA called Transmuter. 
The prefetcher is built on top of Prodigy, the current best-performing data prefetcher for CPUs. 
The key design elements that adapt the prefetcher to the MRA include fused prefetcher status handling registers and a prefetch handshake protocol to support run-time reconfiguration, in addition, a redesign of the cache structure in Transmuter.
An evaluation of popular graph workloads shows that synergistic integration of these architectures outperforms a baseline without prefetcher by 1.27$\times$ on average and by as much as 2.72$\times$ on some workloads.
\vspace{-8pt}

\end{abstract}

\begin{IEEEkeywords}
graph processing, hardware prefetching, reconfigurable architectures, low-power design
\vspace{-7pt}
\end{IEEEkeywords}}

%% file: 01_introduction.tex
\section{Introduction}
With the end of Dennard scaling and Moore's law, traditional architectures like  CPU and GPU do not meet the performance and power requirements of modern data centers and high-performance computing.
Domain-specific accelerators have been proposed to achieve high efficiency, however, their programmability is limited.
Prior works proposed low-power manycore reconfigurable architectures (MRA) to bridge the programmability-efficiency gap.
A recent example is Transmuter (TM)~\cite{transmuter}, which has
demonstrated superior performance on many kernels and  large end-to-end applications~\cite{cosparse,outerspace}. 

Graph analytics is an important domain with many real-world applications~\cite{graph1, graph2}. 
They typically involve irregular memory accesses~\cite{prodigy}, which makes their performance limited on commercial hardware platforms.
Even on a  state-of-the-art (SoTA) MRA such as TM, running five graph analytics workloads (Sec.~\ref{method}) shows an average L1 miss rate of 34\%.
Clearly, this indicates room for performance optimizations.

Several prior works such as~\cite{gapbs, prodigy} propose software (SW) and hardware (HW) optimization techniques to accelerate graph analytics on the CPU. 
Among these, data prefetchers effectively reduce memory access latency by adapting to the complex memory access patterns in graph algorithms. 
Prodigy~\cite{prodigy} is a SoTA prefetcher that has shown the best  performance to date in implementing irregular workloads on general-purpose CPUs.
In this work, we aim to design and adapt a Prodigy-based prefetcher for MRAs and demonstrate its performance improvements on popular graph workloads.

We first discuss why trivially incorporating an unchanged Prodigy into the TM design lends only a small (3\% from our experiments) performance benefit. 
The primary problem is that Prodigy prefetches large quantities of data, and so it cannot be directly used for TM-like MRAs that have
relatively small on-chip caches; they displace hot data that squashes the benefit of bringing in prefetched data.
Furthermore, TM-like MRAs have caches that can be configured in both private and shared modes, whereas Prodigy is not designed to support shared caches.
To adapt Prodigy for TM, we propose the following microarchitectural changes. 
On the prefetcher side, to support shared caching, we propose 1) fusing per-core PreFetch Status Handling Registers (PFHRs) together, 2) a technique to coordinate prefetch requests, and 3) appropriate criteria to squash PF requests. 
On the TM side, we re-evaluate fundamental cache design parameters, including the L1 cache size and the tile-to-L2-bank ratio, in order to better host the prefetched data. 
We evaluate our work on five graph workloads.
We show that our microarchitectural modifications reduce the TM L1 cache miss rate by 40\% on average, which translates to an end-to-end speedup of 1.27$\times$ on average (up to 2.72$\times$) over the baseline TM without prefetching.

In summary, this paper makes the following contributions:
1) presents the challenges in adapting a data-indirect prefetcher, Prodigy, to work with Transmuter, a SoTA MRA; 
2) proposes microarchitectural support to adapt Prodigy for TM to improve the performance on graph workloads; and
3) presents an extensive evaluation showing that the prefetcher-enhanced TM achieves an average of 1.27$\times$ speedup over the baseline TM. 



%% file: 02_background.tex
\vspace{-10pt}
\section{Background and Related Work}
\label{background}

\subsection{Transmuter: A Manycore Reconfigurable Architecture}
Transmuter~\cite{transmuter} is a highly reconfigurable, manycore tiled architecture that optimizes for both programmability and efficiency. 
At the top-most level, TM comprises multiple clusters connected to high-bandwidth memory (HBM) stacks.
Each cluster consists of multiple processing tiles (Fig.~\ref{arch_overview}(a)), that can access the HBM through a reconfigurable L2 cache-crossbar layer. 
It can also communicate via a small but fast scratchpad that is mainly used for storing synchronization variables.
Each  processing tile (Fig.~\ref{arch_overview}(b)) consists of two types of programmable, in-order cores with shallow pipelines, namely the local control processors (LCPs) and the general-purpose processing elements (GPEs). 
The GPEs are the worker cores that receive instructions from the LCP over a work/status queue interface consisting of a set of FIFO buffers that are memory mapped to special addresses. 
The GPEs interface with the L2 through an L1 layer consisting of reconfigurable DCaches (R-DCaches) and reconfigurable crossbars (R-XBars).
This work proposes microarchitectural changes primarily to this L1 layer.

\vspace{-10pt}
\subsection{Prodigy: A SoTA Prefetcher for Irregular Workloads}
Prodigy is an SW-HW co-designed solution to improve the memory system performance of important data-indirect irregular workloads~\cite{prodigy}.
To capture the program behavior, Prodigy proposes a graph representation (different from the input dataset in a graph algorithm) called Data Indirection Graph (DIG).
A DIG is a weighted graph that effectively captures the layout and memory access patterns of key data structures.
Prodigy employs a compiler analysis technique to automatically identify indirect memory accesses and construct the DIG representation into a program's binary.
At run-time, this DIG is used to program a hardware prefetcher to understand an irregular algorithm's program behavior.
A Prodigy prefetcher (PF) engine is placed in the L1 cache of each core.
Each PF engine contains two parts: the \textit{DIG Table} and \textit{PF Logic} (Fig.~\ref{arch_overview}(c)). 
Using small memory elements to store the DIG (DIG tables) and run-time information to keep track of non-blocking live prefetch sequences (PFHRs), the \textit{PF Logic} issues prefetch requests that follow the algorithmic behavior.
It monitors the demand requests from the CPU side, memory fills from the DRAM side, and uses a flexible prefetching algorithm that maintains timeliness by dynamically adapting its prefetch distance to the execution pace of an application.
This work proposes microarchitectural changes primarily to the PFHRs and prefetch handshake protocol.

 
\vspace{-10pt}
\subsection{Related Work}

Prior work proposes MRAs that can reconfigure at the sub-core level~\cite{subcorereconfig} and network-level~\cite{simba}.
Transmuter~\cite{transmuter} uses general-purpose cores and reconfigures its on-chip memory and interconnect, providing programmability and flexibility. 
CoSPARSE~\cite{cosparse}, which is orthogonal to this work, optimizes the performance of graph analytics on TM by adapting between pull-push mode of graph algorithms.
There are many works on prefetching for improving the memory access latency of graph analytics. Typically, they use knowledge of how graph data is laid out in memory to issue prefetch requests~\cite{ainsworth_graph, droplet, prodigy}.


\input{mainFig}

%% file: mainFig.tex
\begin{figure}
    \centering
    \includegraphics[width=1\columnwidth]{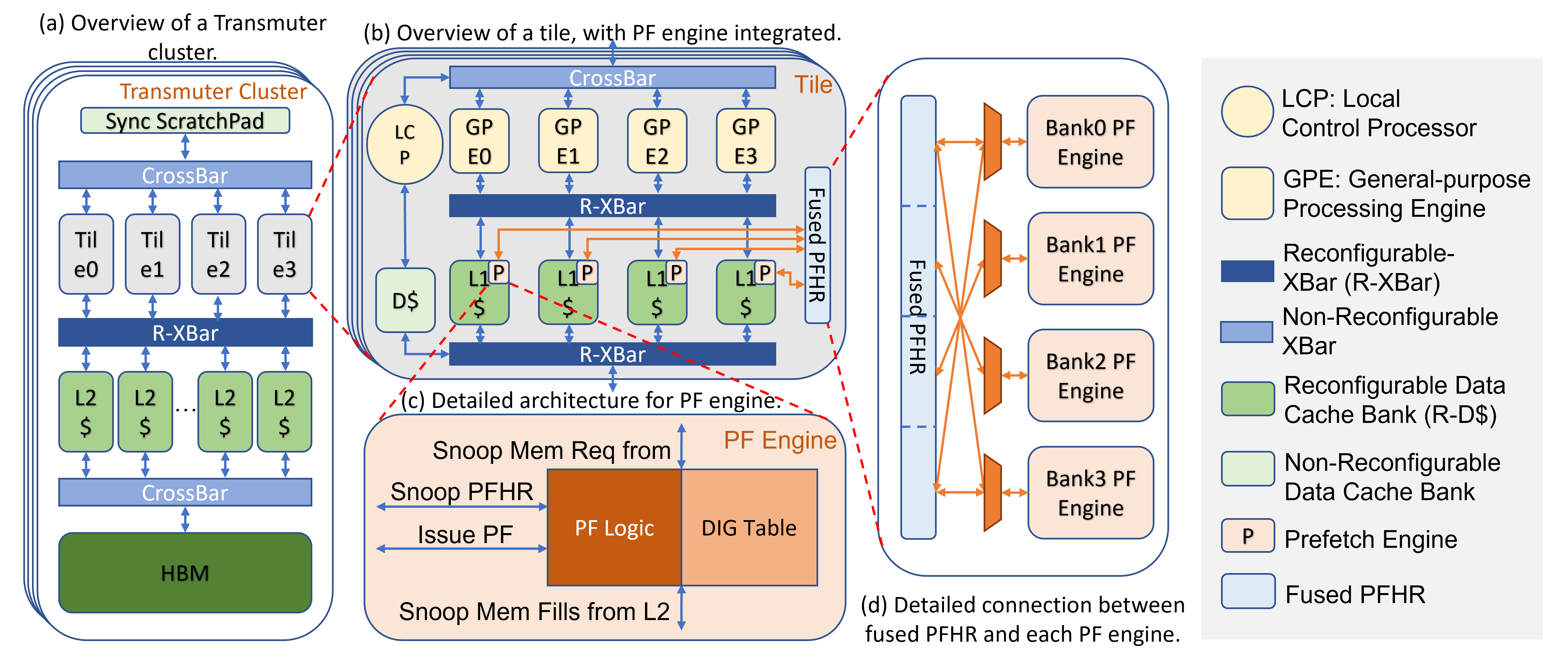}
    \vspace{-15pt}
    \caption{(a) Overview of a TM cluster with four tiles. No changes are introduced in this level. (b) Overview of a Transmuter tile with four GPEs per tile. The prefetch engine is attached to each L1 R-DCache bank. (c) Detailed architecture of a PF engine. 
    (d) Detailed connection between PF engines within a tile and the fused PFHR. } 
    \label{arch_overview}
    \vspace{-15pt}
\end{figure}

%% file: 03_arch.tex
\vspace{-13pt}
\section{Adapting Prodigy for Transmuter}
\label{arch}



\subsection{Prodigy Microarchitectural Augmentation}
\textbf{Challenges: }
Experiments show that graph algorithms generally perform better in shared mode (Sec~\ref{shared_vs_private}).
However, Prodigy is originally designed for a CPU architecture with a private L1 cache. 
The original PF engine only monitors the prefetch requests and prefetches fill into its own cache without communicating with the other PF engines. 
MRAs such as TM support shared L1 caching and even provide run-time reconfiguration capability, such as switching between private/shared cache modes.
While the private mode is similar to a CPU architecture where a PF engine in each cache bank only monitors its own traffic, the shared mode is different.
In the shared mode, data is stored in the cache banks using a cache coloring technique and all PF engines within a tile must coordinate with each other to generate requests by the correct prefetcher and bring data into the correct cache bank.
With unchanged Prodigy, we observe that the prefetched data is brought into a bank to which the address does not map. 
Thus, the prefetched data is unused and causes cache pollution, resulting in minimal performance gains of 3\% over a no-PF baseline.
Next, we discuss \textbf{Prodigy microarchitectural changes}. 


\vspace{-5pt}
\subsubsection{Fused PFHR Array}
In order to support TM's run-time cache reconfiguration, the PFHR needs to incorporate reconfigurability.
In particular, since the PFHR stores the applications' run-time information, it should be private to a single PF engine, or shared among all the PF engines within a tile, depending upon the cache configuration. 
As shown in Fig.~\ref{arch_overview}(d), we propose a \textbf{fused PFHR array} within a tile, instead of one PFHR per PF engine as in the original design~\cite{prodigy}. 
The fused PFHR array is managed in banks. Based on the cache mode, a PF engine will access the corresponding bank(s) of the fused PFHR array and use CAM to search for run-time information.
Specifically, when the L1 cache is private, each PF engine accesses its own dedicated PFHR bank, and when the L1 is shared, all the PF engines within atile access all banks of the PFHR array. 
Furthermore, for fairness and design simplicity, one PFHR bank has only one read/write port, and the PF engine can access the fused PFHR array in a round-robin fashion in shared mode. 
Experiments show that this round-robin arbitration has a negligible performance impact compared to unlimited read/write ports. 
With this microarchitectural enhancement, all PF engines can access their dedicated PFHR bank in parallel in the private cache mode, whereas one PF engine has access to the whole PFHR array in the shared cache mode, per clock cycle. 

\vspace{-5pt}
\subsubsection{Prefetch Request Handshake Protocol}
When the L1 caches are shared, a cache coloring technique, similar to page coloring~\cite{pageColor}, is applied to the L1 cache banks to improve cache utilization and minimize bank contention. 
However, when a prefetcher is involved in the design, it generates prefetch requests to bring data into its host cache bank, which may differ from the bank that the prefetched data is mapped to, \textit{i.e.} where it should be fetched into, dictated by the cache coloring policy. 
This violates the cache coloring policy and thus prevents the cores from accessing the prefetched data. 
Therefore, we propose a prefetch request handshake protocol among the PF engines within a tile.
This technique guarantees that each prefetch request will be issued by the PF engine corresponding to the requested addresses' host cache bank(s). 
Our policy stalls the PF engine from immediately issuing a generated prefetch request. 
Instead, it follows the cache coloring policy and passes the prefetch requests to the corresponding cache bank's PF engine and lets that prefetcher issue the prefetch requests. 

\vspace{-5pt}
\subsubsection{Redesigned PFHR Squash Criteria}
The PFHRs in Prodigy uses a squashing mechanism to keep the prefetcher always running ahead of the program execution. 
When the execution thread catches up with the prefetcher, new run-time information overwrites the old one. 
However, when the L1 cache is shared, the PFHR array is shared between all the PF engines. 
Since the progress of different GPE cores may not be the same, a PFHR entry generated by a specific GPE should not be overwritten by other GPEs within the same tile. 
In order to correctly squash the PFHR entries, we add the GPE-ID as an additional field to the PFHR entry and enforce that only matching GPE-ID entries can be squashed. 

\vspace{-10pt}
\subsection{Transmuter Cache Design Changes}
\label{sec:tm_adap}
\textbf{Challenges: }
While CPUs have multiple levels of large caches (up to several megabytes) that offer enough storage for hosting large amounts of prefetched data, GPUs, as well as low-power MRAs such as TM, have far smaller on-chip storage to compute the ratio.  
Therefore, directly integrating Prodigy, which is optimized for multicore CPUs, with TM leads to high L1 cache replacements and congestion at the L1-to-L2 interface. 
We carefully redesign the cache structure in TM in order to ease this L1-to-L2 bottleneck. 
Specifically, we make the following changes to \textbf{the TM architecture} (detailed evaluation in Sec.~\ref{sec:designspace}).

\vspace{-5pt}
\subsubsection{Resizing the L1 Caches}
Using a Prodigy-based prefetcher generates a large number of prefetch requests ahead of the program execution, which need to be stored in the L1 cache for future use. 
Based on our experiments (Sec.~\ref{sec:cache_size}), we notice a high replacement count (number of evictions of a valid block when bringing data into the cache) for the L1, which indicates that it is too small to hold the additional prefetched data.
One immediate solution is to tune down the prefetcher aggressiveness, but we observed that it significantly degrades the performance improvement (less than 5\% speedup). 
Our evaluation with CACTI shows that access latency is 1 cycle for SRAM smaller than 64kB at 1GHz clock frequency. Considering the performance gain from resizing, we decide to increase the L1 cache size from 4kB to 16kB per bank to create more space for prefetched data, while slightly increasing the energy and area. 

\vspace{-5pt}
\subsubsection{Finer-Grained Banking in the L2}
A high number of requests generated by the prefetcher also leads to the congestion at L1-to-L2 interface. 
As the R-XBars serialize requests destined for the same output port, with more prefetch requests generated, the requests are kept waiting at local buffers to be forwarded (Sec.~\ref{sec:banks}).
This is exacerbated by the fact that the original TM architecture assigns the L2 bank count to be equal to the number of tiles, \textit{i.e.} it does not consider the increased pressure on the L2 banks when there are too many L1 banks per tile.
We tackle this problem by increasing the banking granularity at the L2.
Specifically, increase the number of L2 banks per tile from 1 to 4 (\textit{i.e.} 16 L2 cache banks in total, for a TM design with four tiles), \textit{while keeping the total L2 cache size the same}.
This improves the L1-L2 connection bandwidth without much additional area or energy overhead. 



%% file: 04_method.tex
\vspace{-12pt}
\section{Methodology}
\label{method}

\subsection{Implementation of Graph Algorithms on Transmuter}

As there are no open-source graph analytics implementations for TM, we developed hand-optimized code for five algorithms using the TM C++ API. 
These include PageRank (PR), PageRank-Nibble (PRN), Breadth-First Search (BFS), Single-Source Shortest Path (SSSP), and Collaborative Filtering (CF).
These implementations operate using a pull mode, iterating over each node's incoming vertices to see if a self-update is needed, and require a compressed sparse column format.

\vspace{-5pt}
\subsection{Experimental Setup}

\begin{table}
\caption{Microarchitectural parameters of the gem5 model.}
\vspace{-5pt}
\resizebox{\columnwidth}{!}
{
\begin{tabular}{|m{2.7cm}||m{9cm}|}
\hline
\multicolumn{1}{|c||}{\textbf{Module}} & \multicolumn{1}{c|}{\textbf{Microarchitectural Parameters}}  \\ \hline\hline
{PE/LCP}         & 1-issue, 4-stage, in-order (\texttt{MinorCPU}) core @ 1.0 GHz
\\ \hline
{L1 RCache (per bank)} & 16~kB, 1-ported, word-granular, 4-way set-associative non-coherent cache with 8~MSHRs and 64~B block size, 1 bank per GPE
\\ 
\hline
{L2 Cache (per bank)} &  4~kB, shared mode, others are the same as L1 R-DCache, 4 banks per tile
\\ \hline
{Main Memory}         & 1 HBM2 stack: 16 64-bit pseudo-channels, each @ 8000 MB/s, 80-150~ns average access latency
\\ \hline
Prefetcher PFHR & 8 entries per GPE
\\ \hline
\end{tabular}
} 
\vspace{-15pt}
\label{tab:gem5params}
\end{table}

We use gem5~\cite{gem5} to simulate the enhanced Prodigy-TM system. 
The microarchitectural parameters used are similar to prior works~\cite{transmuter, cosparse, prodigy} and are listed in Tab.~\ref{tab:gem5params}. 
We focus on the 4$\times$16 TM configuration (4 tiles with 16 GPEs/tile) with our modifications from Sec.~\ref{sec:tm_adap}.
This design comprises one 16kB L1 cache bank per GPE and four 4kB L2 cache banks per tile. 
The L1 and L2 caches are in the shared mode (\textit{i.e.} L1 banks shared across GPEs within a tile; L2 banks shared across tiles), unless otherwise stated.
The original TM design without a prefetcher acts as the baseline to evaluate the performance of the Prodigy-based prefetcher using our five graph application implementations. 
Tab.~\ref{tab:graphs} lists the details of each of the real-world~\cite{snapnets, florida_matrix}, uniform-random~\cite{nx} and Kronecker graphs~\cite{kronecker} used in the experiments. 
Following the prior work~\cite{transmuter, cosparse}, we use CACTI~7.0~\cite{cacti} for our energy estimations.
We build a model of the whole system, including the redesigned cache structure, and cross-verify it with a fabricated chip prototype~\cite{kim2022versa}.


%% file: 05_result.tex
\vspace{-10pt}
\section{Evaluation}
\label{result}
\subsection{Overall Performance}

\begin{figure}
    \centering
    \vspace{-10pt}
    \includegraphics[width=0.85\columnwidth]{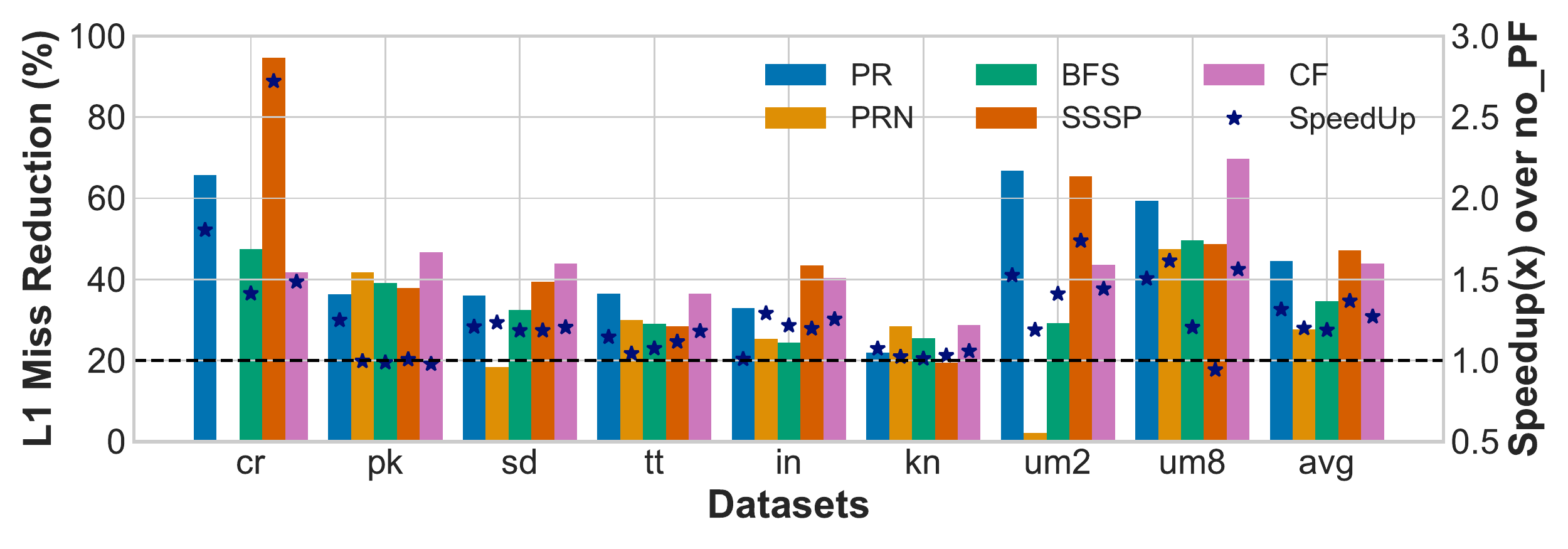}
    \vspace{-10pt}
    \caption{L1 miss reduction (bars) and speedup (markers) achieved by the proposed system over the baseline 4$\times$16 TM. The best prefetcher aggressiveness is set for each experiment. CARoad-PRN is not shown because it exceeded the simulation limit.}
    \label{fig:overall}
    \vspace{-15pt}
\end{figure}

Fig.~\ref{fig:overall} shows the reduction in L1 cache miss rate and overall speedup achieved by the use of our Prodigy-Transmuter design, across the evaluated different applications. 
The proposed system achieves an average of 1.27$\times$ better performance compared to the baseline TM design.
These gains are a direct consequence of 84\% prefetch accuracy and 40\% L1 cache miss rate reduction on average.
We, thus, conclude that a combination of the Prodigy microarchitectural changes and redesigned TM cache structure provides promising performance benefits.

We observe that the system generally favors graphs that are sparse and more uniformly distributed, such as \texttt{cr}, which shows the highest speedup (2.72$\times$) for SSSP.
For such sparse graphs, given the constant cache size, the prefetcher can run further ahead of the application (compared to denser graphs) to prefetch more data, thus better hiding the memory access latency. 
For non-uniform graphs, when encountering a highly-connected node, the prefetcher brings all its neighbors' data into the caches, which evicts prior prefetched data that was about to be used.
Thus, the memory access latency for these evicted data increases, which affects the overall performance. 

We emphasize that our reported improvements are over the TM baseline that already provides superior performance gains over commodity hardware for the domains of sparse algebra and graph analytics~\cite{transmuter,cosparse}. 
Our proposal, thus, leads to an estimated 4.6$\times$ improvement over the CPU (i7-6700K running MKL library) with a 4$\times$16 TM on the same workloads.

Due to space limitations, we only report a qualitative comparison of Prodigy with other prefetchers. Prodigy can catch two general data access patterns that are prevalent in graph analytics and can issue prefetches with awareness of the application progress~\cite{prodigy}. Less sophisticated prefetchers such as stride and correlating prefetchers provide limited effectiveness for irregular memory access patterns since they are unaware of application run-time information.
\input{tab2}

\vspace{-10pt}
\subsection{Design Space Exploration} \label{sec:designspace}
We conduct a thorough design space exploration to select the optimal prefetcher-enhanced TM configuration. 

\vspace{-5pt}
\subsubsection{Private versus Shared L1 Cache} 
\label{shared_vs_private}
Even though data-dependent memory accesses form the dominant pattern in graph algorithms, temporal locality still exists, especially for power-law graphs. 
Nodes with higher degrees tend to be revisited more often. 
The shared cache exploits this locality better than the private cache. 
Running PR without a prefetcher, the shared L1 system performs, on average, 1.51$\times$ better than the private L1 system, with up to 2.68$\times$ speedup for the Slashdot dataset.
Even with the prefetcher enabled in both modes, the shared mode performs 1.33$\times$ better than the private mode.
Thus, the shared L1 cache generally performs better.

\begin{figure}
     \centering
     \begin{subfigure}{0.59\linewidth}
         \centering
         \includegraphics[width=0.85\columnwidth]{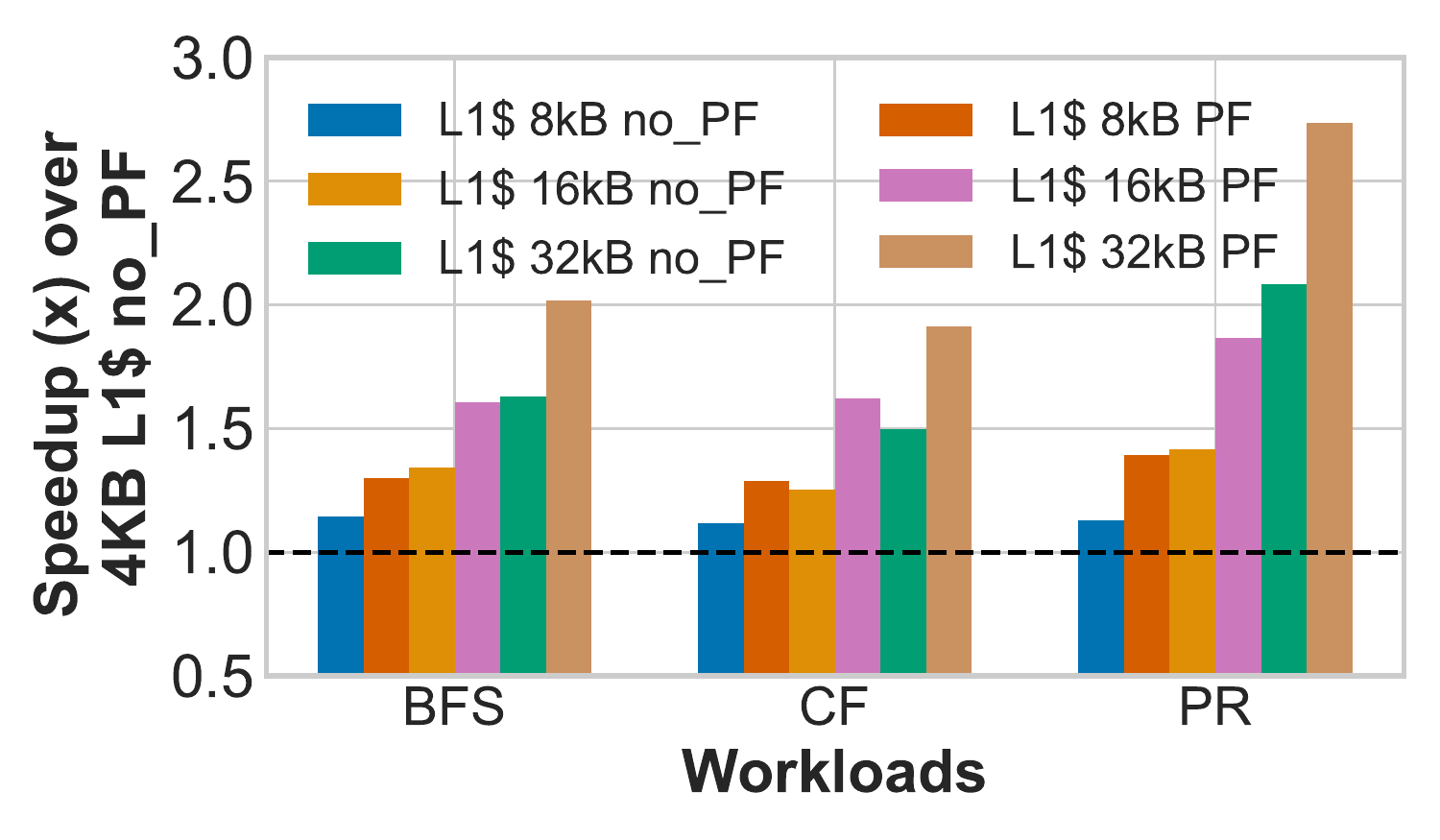}
         \vspace{-10pt}
         \label{fig:cache_size_a}
     \end{subfigure}
     \hfill
     \begin{subfigure}{0.39\linewidth}
         \centering
         \includegraphics[width=0.85\columnwidth]{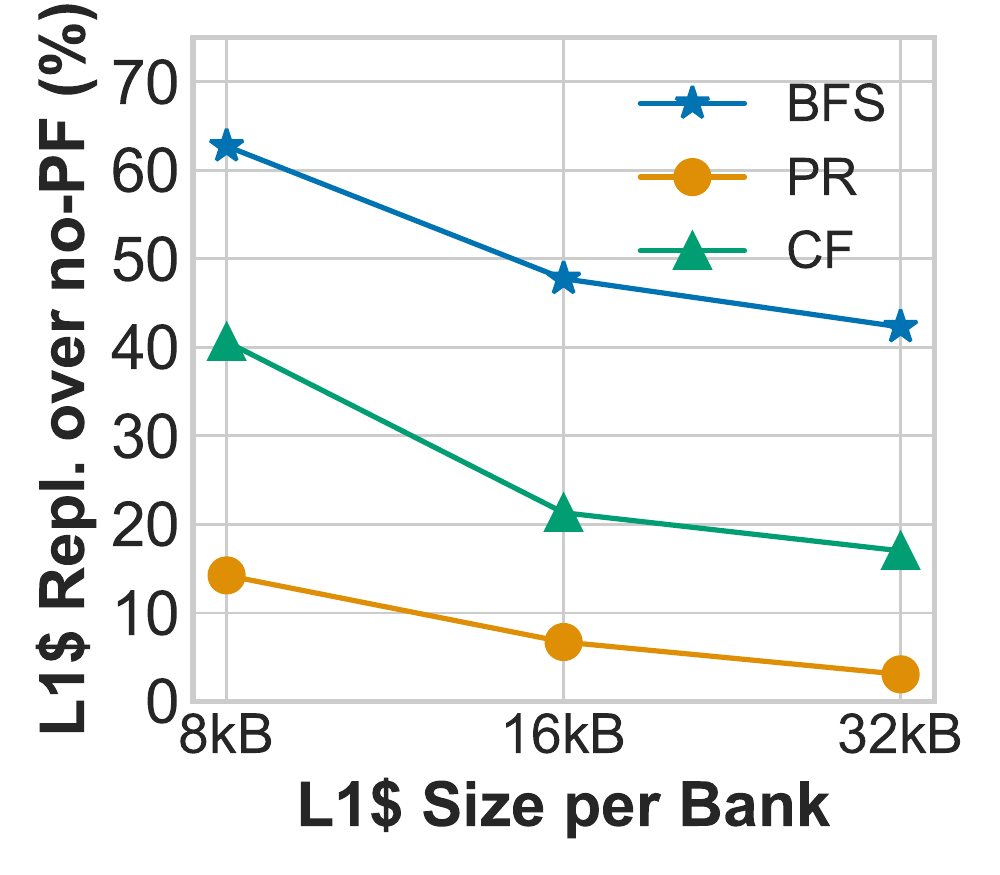}
         \vspace{-10pt}
         \label{fig:cache_size_b}
     \end{subfigure}
    \caption{L1\$ size experiment. Left: Speedup over TM with 4KB L1\$ per bank no-PF, for different L1\$ sizes with and without PF. Right: \% of additional L1 cache replacements over no-PF configuration. Datasets used: cr, pk, sd, tt, in, um2, and um8.}
    \label{fig:cache_size}
    \vspace{-10pt}
\end{figure}

\vspace{-5pt}
\subsubsection{L1 Cache Sizes}\label{sec:cache_size}
Fig.~\ref{fig:cache_size} (left) shows the performance gains across different L1 cache sizes, with and without the PF.
The PF works better with a larger L1 cache because the system can hold more prefetched data, which effectively hides the memory access latency. 
The benefits tend to saturate (after 32kB per bank, in our case). 
We further observe from Fig.~\ref{fig:cache_size} (right) that the L1 cache replacement count is much higher than the no-PF configuration, and gets better with an increase in L1 cache capacity. 
As the cache access latency is still 1 cycle, the overall performance improves (for up to 32kB) when the L1 cache size is increased.
The 16kB-PF configuration increases 20\% in area (original 64$\times$64 TM design is only 1.7\% of a CPU, i7-6700K, area~\cite{transmuter}) and 22\% in energy compared to the 4kB-noPF design. But the system can achieve 1.68$\times$ speedup. 
Comparing with 4kB-noPF in terms of EDP, we observe reductions of +2\%, +15\%, +22\% and -2\% for the designs with 4kB-PF, 8kB-PF, 16kB-PF and 32kB-PF, respectively.
Thus, we select the design with 16kB per bank L1 cache to trade area and energy for better performance.

\vspace{-5pt}
\subsubsection{Number of L2 Cache Banks}\label{sec:banks}
\begin{figure}
\vspace{-10pt}
     \centering
     \begin{subfigure}{0.59\linewidth}
         \centering
         \includegraphics[width=0.85\columnwidth]{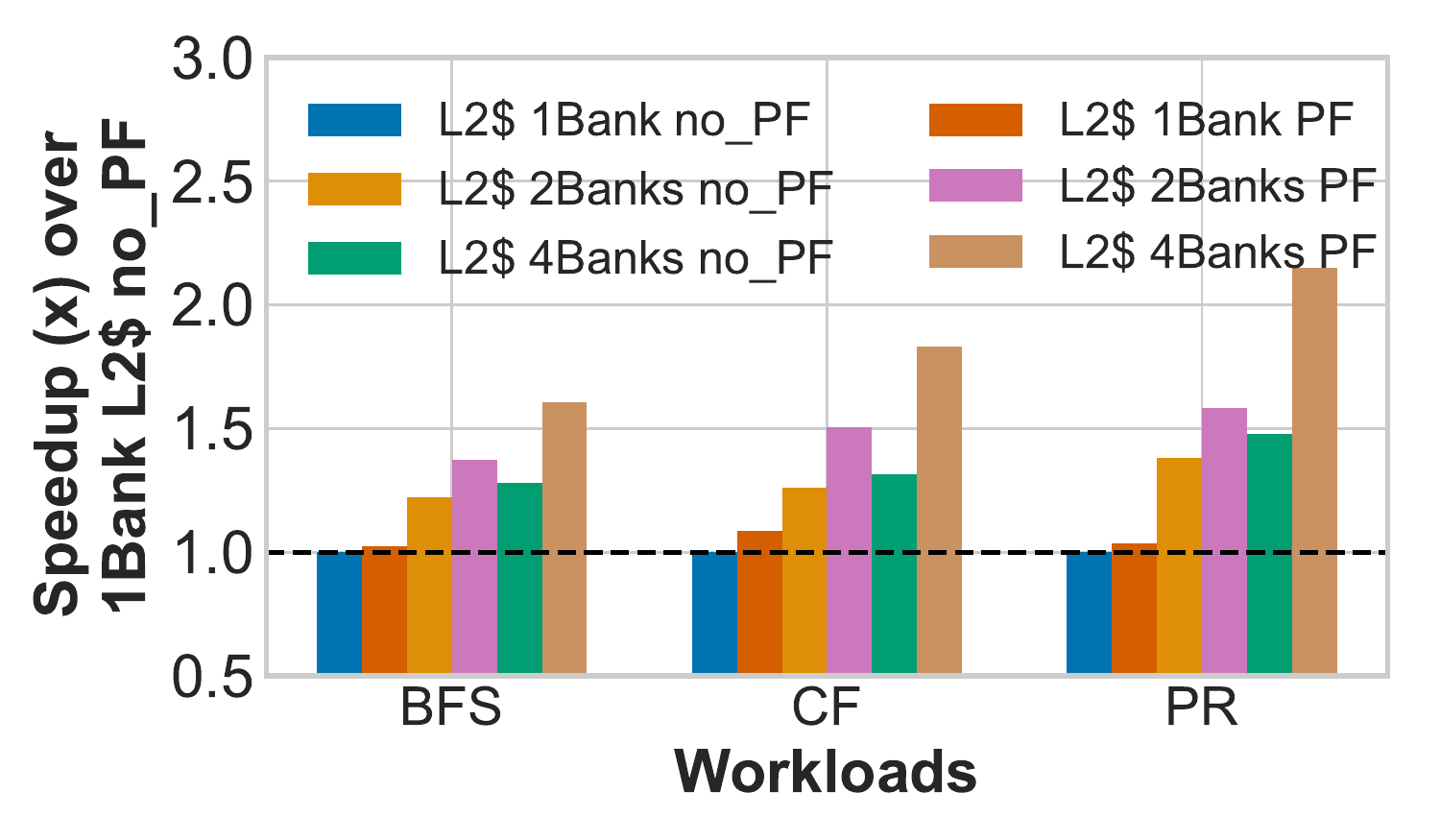}
         \vspace{-10pt}
         \label{fig:L2Bank1}
     \end{subfigure}
     \hfill
     \begin{subfigure}{0.39\linewidth}
         \centering
         \includegraphics[width=0.85\columnwidth]{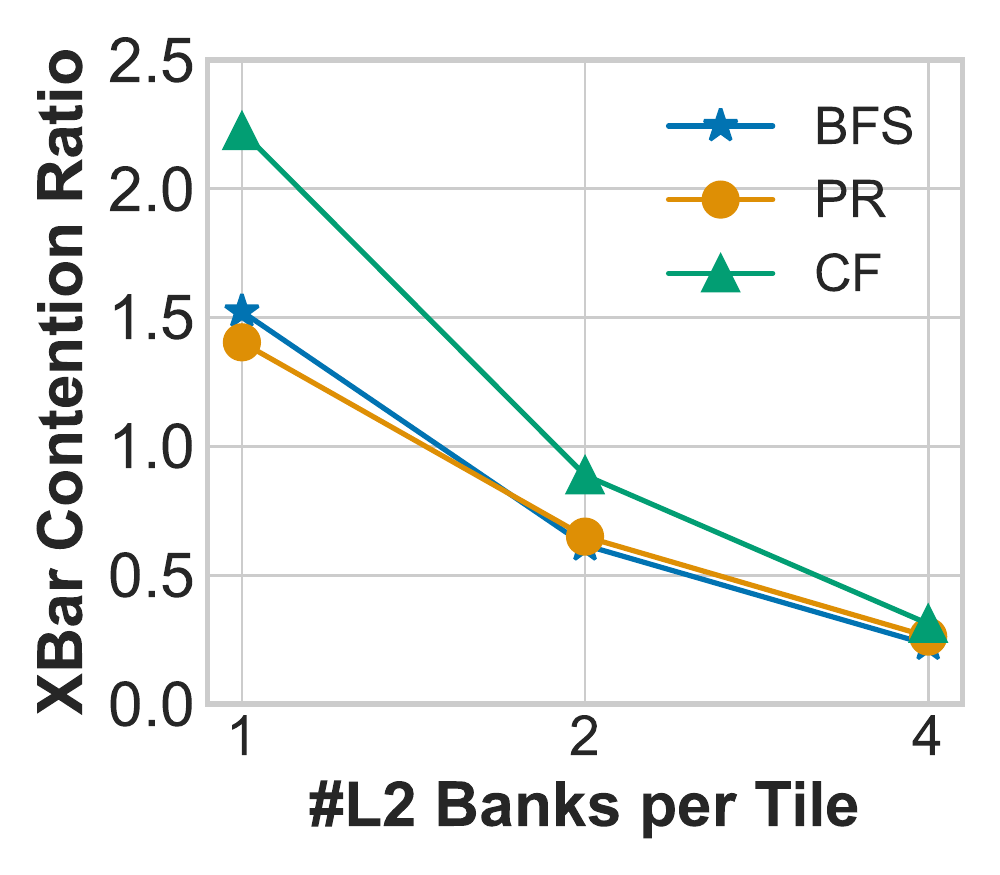}
         \vspace{-10pt}
         \label{fig:L2Bank2}
     \end{subfigure}
    \caption{The number of L2\$ banks experiment. Left: Speedup over one L2 bank per tile no-PF, for different numbers of L2\$ banks per tile with and without PF. Right: Contention ratio at L1-to-L2 R-XBar for different numbers of L2\$ banks per tile, PF enabled. Datasets used: cr, sd, tt, um2, and um8.}
    \label{fig:L2Bank}
    \vspace{-10pt}
\end{figure}

The use of the Prodigy-based prefetcher results in more memory requests generated from the CPU side. 
We experiment with different numbers of L2 cache banks per tile while keeping the total L2 cache size the same. 
Fig.~\ref{fig:L2Bank} (left) shows the performance gains with additional L2 cache banks and by enabling the prefetcher, and Fig.~\ref{fig:L2Bank} (right) shows the contention ration at the L1-to-L2 R-XBar for the prefetcher enabled configuration. 
The contention ratio is obtained from the simulator, for a number of cycles, dividing the total packets queued at R-XBar by the total packets that go through the R-XBar, and then averaged across the workload execution. 
A higher contention ratio implies that the input ports are more congested with requests.
When there is only one L2 cache bank per tile, the memory requests are bottlenecked by the processing speed of the R-XBar (higher contention ratio). 
When more L2 cache banks are added, the R-XBar can process more requests at the same cycle, increasing the L1-L2 bandwidth, and thus improving the performance. 
Additional L2 banks also decrease the probability of bank conflicts at the L2, which further reduces the contention ratio.
However, adding more L2 cache banks leads to performance saturation at two banks per tile. Only with prefetcher enabled can the system  make full use of the larger cache bandwidth.
Therefore, we configure the L2 caches to have 4 banks per tile and 16 banks in total for the 4$\times$16 TM. 

\vspace{-15pt}
\subsection{Overhead Analysis}
\subsubsection{Area and Energy Overheads}
Using the model from previous work~\cite{prodigy}, Prodigy introduces a negligible 0.28kB storage overhead per GPE (PFHR and DIG storage) in area. 
The energy overhead of Prodigy (PFHR and DIG storage) is estimated using a fully-associative cache with block size and the number of entries matched with that of the PFHR/DIG. 
On average, the overhead of the prefetcher is only 3.42\% of the total system energy, which is negligible. 
If considering the L1 size increase as area overhead, the final design increased 20\% comparing to the original 4kB-L1 Transmuter.

\vspace{-5pt}
\subsubsection{Performance and Efficiency for Different Configurations}
\begin{figure}
    \centering
    \vspace{-10pt}
    \includegraphics[width=0.8\columnwidth]{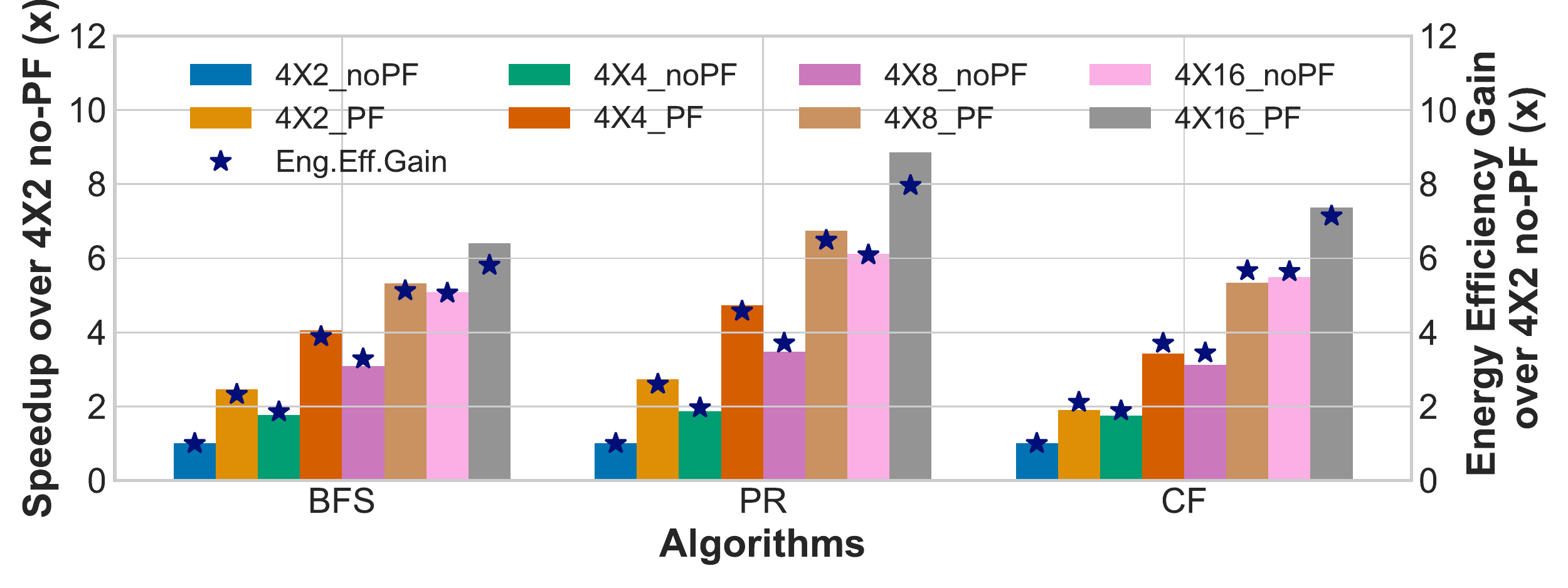}
    \vspace{-10pt}
    \caption{Speedup (bars) and energy efficiency gain (markers) achieved by the proposed system. The data is normalized to the execution time of the 4$\times$2 TM with no prefetcher.}
    \label{fig:tmsize}
    \vspace{-15pt}
\end{figure}

Fig.~\ref{fig:tmsize} shows the speedups and energy efficiency gains for different TM configurations. 
The total L1 and L2 cache sizes are kept constant across the experiments. 
The workloads' performance scales well with the TM dimensions. 
We further observe that the smaller TM with prefetcher has a better performance compared to the larger TM without prefetcher in terms of both time and energy efficiency (1.15$\times$ speedup on average). 
Therefore, we conclude that a smaller architecture with the Prodigy prefetcher would be much more efficient for graph analytics compared to a larger architecture without the prefetcher.

%% file: tab2.tex
\begin{table}
 \vspace{0pt}
\caption{Specification for the input graphs. \texttt{MemSize} is calculated based on memory footprint with the PR algorithm.}
\vspace{-5pt}
\centering
\resizebox{0.85\columnwidth}{!}
{
\begin{tabular}{|c||r|r|r||c|}
\hline
\textbf{Graphs} & \multicolumn{1}{c|}{\textbf{\# Vertices}} & \multicolumn{1}{c|}{\textbf{\# Edges}} & \multicolumn{1}{c||}{\textbf{MemSize}} & \textbf{Type} 
\\ \hline \hline
CARoad(cr) &	1,965,206 &	2,766,607& 40.540 MB & Road Net 
\\ \hline 
soc-Pokec(pk) &	1,632,803 &	30,622,564 & 141.730 MB& Social Net
\\ \hline
Slashdot0811(sd) &	77,360 &	905,468 & 4.635 MB& Social Net
\\ \hline
ego-Twitter(tt) &	81,306 &	1,768,149 & 7.986 MB& Social Net
\\ \hline
in-2004(in)	 & 1,382,908 &	16,917,053 & 85.635 MB& Web Net
\\ \hline\hline
Kronecker18(kn) &262,143&	3,805,448	&	18.517 MB& Syn Graph 
\\ \hline
Uni\_gr\_1Mx2(um2) &	1,000,000&	2,000,000  & 22.888 MB& Syn Graph 
\\ \hline
Uni\_gr\_1Mx8(um8) &	1,000,000&	8,000,000  & 45.776 MB& Syn Graph 
\\ \hline
\end{tabular}
\vspace{-20pt}} 
\label{tab:graphs}
\end{table}

%% file: 07_conclusion.tex
\vspace{-10pt}
\section{Conclusion}
\label{conclusion}

In this paper, we presented the design of an irregular workload prefetcher for Transmuter (TM), a manycore reconfigurable architecture (MRA). 
We motivated why a na\"ive incorporation of a prior prefetcher would not suffice to bring improvements into the MRA, as opposed to the multicore CPU system it was designed for.
We presented a set of microarchitectural changes in the TM with the Prodigy that works with the run-time cache reconfiguration of the MRA.
We further redesigned the cache structure of TM to store more prefetched data and alleviate congestion at the L1-to-L2 interface.
Our simulations show that the proposed system achieves an average speedup of 1.27$\times$ over the baseline system without a prefetcher. 
Due to space limitations, this work only demonstrates our approach for TM. However, it can be readily extended to adapt generic prefetchers onto MRAs.